\documentclass[sigconf]{acmart}
\usepackage{threeparttable}

\AtBeginDocument{%
  \providecommand\BibTeX{{%
    \normalfont B\kern-0.5em{\scshape i\kern-0.25em b}\kern-0.8em\TeX}}}

\setcopyright{acmcopyright}
\copyrightyear{2023}
\acmYear{2023}
\acmDOI{XXXXXXX.XXXXXXX}

\acmConference[ECIR '23]{European Conference on Information Retrieval}{April 02-06, 2023}{Dublin, Irland}
\begin{document}

\title{Opening the TAR Black Box: Developing an Interpretable System for eDiscovery Using the Fuzzy ARTMAP Neural Network}

\author{Charles Courchaine}
\orcid{0000-0002-5404-9438}
\affiliation{%
  \institution{National University}
  \country{United States}
}
\email{charles@courchaine.dev}

\author{Ricky J. Sethi}
\orcid{0000-0001-5254-3750}
\affiliation{%
  \institution{Fitchburg State University}
  \institution{National University}
  \country{United States}
}
\email{rickys@sethi.org}
\renewcommand{\shortauthors}{Courchaine and Sethi}




\maketitle

\section{Introduction}
Technology-assisted review (TAR) utilizes an information retrieval system to discover all, or nearly all, the relevant documents in a corpus and help reduce the human effort required to find these documents \cite{YangEtAl2019, CormackGrossman2015, ChhatwalEtAl2018}. TAR systems are employed in high-recall tasks such as e-discovery, systematic literature reviews, evidence-based medicine, and information test collection annotation \cite{YangEtAl2019, CormackGrossman2015}. These systems often employ a document classifier and an active learning component to select what documents a human should review \cite{ChhatwalEtAl2017, YangEtAl2021a}. 
A TAR system that can explain how and why document relevance predictions are made is a vital tool for enabling attorneys to meet their ethical obligations to clients and enable clients to fully participate in the process \cite{Endo2018}. Despite the benefits of an explainable TAR system, current systems fail to deliver on why documents are classified as responsive and so these systems are still typically perceived as “black boxes” by practitioners \cite{ChhatwalEtAl2018, MahoneyEtAl2019}.

While a few studies have attempted to bring explainability to TAR systems, they focused on extracting snippets from the documents as the mechanism of explanation rather than directly explaining the relevance model \cite{ChhatwalEtAl2018, MahoneyEtAl2019}. Instead, we looked at the explainable Fuzzy ARTMAP algorithm. The model learned by the Fuzzy ARTMAP algorithm can be directly interpreted geometrically \cite{CarpenterEtAl1992, MengEtAl2019} or as a set of fuzzy If-Then rules \cite{CarpenterTan1993, CarpenterTan1995}, depending on the features used. 

We performed an initial evaluation of the performance of the explainable Fuzzy ARTMAP algorithm in the TAR domain and found robust performance in terms of recall and precision \cite{CourchaineSethi2022}. Building on the strength of these initial results, we have now continued this foundational research by:
\begin{itemize}
  \item performing a hyperparameter sweep to refine the parameters
  \item evaluating the system against the 20Newsgroups, Reuters-21578, RCV1-v2, and Jeb Bush emails corpora for recall, precision, and F$_{1}$, and 
  \item generating If-Then rules of document relevance
\end{itemize}

While these corpora are not specific to the legal domain, the RCV1-v2 and Jeb Bush emails corpora are frequently used in e-discovery evaluations \cite{YangEtAl2019, YangEtAl2021} because legal matters are often confidential \cite{ChhatwalEtAl2018, CormackGrossman2015} and their corpora are unavailable.  The 20Newsgroups corpus is commonly used as a test corpus with ART-based algorithms \cite{MengEtAl2019, MarcekRojcek2017}; it and the Reuters-21578 corpus are also commonly used in evaluating text classification algorithms \cite{AltinelGaniz2018}.

\section{Fuzzy ARTMAP}
Adaptive Resonance Theory (ART) describes how the brain learns and predicts in a non-stationary world \cite{Grossberg2021}. This theory models how brains can quickly learn new information without forgetting previously learned information. ART has been implemented in numerous neural network architectures for supervised, unsupervised, and reinforcement learning applications \cite{BritodaSilvaEtAl2019}.
Fuzzy ART is a neural network algorithmic instantiation of ART that utilizes operators from fuzzy set theory; specifically, the fuzzy AND operator, to work with real-valued features \cite{CarpenterEtAl1992}. The supervised version of the Fuzzy ART algorithm is the Fuzzy ARTMAP algorithm that maps between inputs and categories. By integrating fuzzy set theory and ART dynamics in the Fuzzy ARTMAP neural network algorithm, various interpretations of the learned model are possible. What the model learns may then be represented as fuzzy If-Then text-based rules or depicted geometrically \cite{CarpenterEtAl1992, Grossberg2020}.

\begin{figure}[b]
  \centering
  \includegraphics[width=0.53\columnwidth]{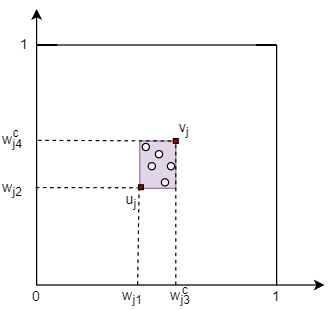}
  \caption{With complement encoding and a 2-dimensional input, the \textit{j}$^{th}$ category represented by weight vector \textit{w} can be interpreted geometrically as a rectangle with corners u$_{j}$ and v$_{j}$, with u$_{j}$ corresponding to the first and second positions of the vector, and v$_{j}$ corresponding to the complement encoded third and fourth positions. The circles inside the rectangle indicate inputs that fall within the category bounds.}
\label{fig:fuzzy_geometry}
\end{figure}

To take advantage of the geometric interpretation, however, the input must be complement encoded. Complement encoding is a normalization method when working with Fuzzy ARTMAP \cite{CarpenterEtAl1992} in which the input vector $\boldsymbol{x}$ is concatenated with its complement $\boldsymbol{\overline{x}}$, yielding an input of $\boldsymbol{I = [x, \overline{x}]}$. As a result, the categories learned by the Fuzzy ARTMAP algorithm can be interpreted as \textit{n}-dimensional hyper-rectangles \cite{CarpenterEtAl1992, MengEtAl2019}. When interpreting the model geometrically, the learned weights from the first half of the vector, the non-complement encoded portion, form one corner of the hyper-rectangle, and the second half of the vector, the complement-encoded portion, forms the other corner as illustrated in Figure \ref{fig:fuzzy_geometry}.

\section{Method}
For the 20Newsgroups, Reuters-21578, RCV1-v2, and Jeb Bush emails corpora, we used tf-idf features with the smaller corpora and the 300-dimension versions of the  GloVe and Word2Vec vectorizations with all of the corpora. 
All the topics in 20Newsgroups, 120 topics in Reuters-21578, and 30 topics in both the RCV1-v2 and the Jeb Bush emails corpora were used for evaluation; the RCV1-v2 and the Jeb Bush corpora were down-sampled to 20\% and 50\% per \cite{YangEtAl2021} due to memory constraints, retaining the general prevalence per topic. For each topic, the Fuzzy ARTMAP algorithm was trained with ten relevant documents and 90 non-relevant documents regardless of corpora size, and the review was run with batches of 100 for the smaller corpora and 1,000 for the larger corpora. The review of documents for each topic concluded when the algorithm predicted no more relevant documents in the unevaluated portion of the corpus.  The Fuzzy ARTMAP algorithm was modified to report the degree of fuzzy subsethood \cite{CarpenterEtAl1992, Kosko1986} associated with documents predicted as relevant, and this degree of fuzzy subsethood was then used to rank the documents for active learning. Based on the results of a sweep of the Fuzzy ARTMAP neural network algorithm hyperparameters, which evaluated different combinations of vigilance ($\rho$) and learning rates ($\beta$), vigilance was set to .95, and a fast learning rate of 1.0 was selected.

A proof-of-concept of one of these If-Then rules for the tf-idf vectorization was produced for predicting documents belonging to the pc.hardware category of the 20Newsgroups dataset, reproduced in Table \ref{tab:rule_example}. The tf-idf feature is in italics, and the level of prevalence is in bold. For this example, the level of prevalence was quantized into three levels: rarely, somewhat, and highly prevalent. 
Additionally, an example of the geometric interpretation is shown and discussed in Figure \ref{fig:fuzzy_geometry}.

\section{Results and Discussion}
Considering all corpora and vectorizations, the Fuzzy ARTMAP-based system achieved 100\% recall 31\% of the time, and achieved the suggested floor of 75\% \cite{KeelingEtAl2020} or better recall 67\% of the time, as seen for median recall, precision, and F$_{1}$ in Table \ref{tab:median_metrics}. Recall between the vectorizers for the Reuters-21578 and 20Newsgroups corpora was different by a statistically significant degree based on a Friedman test \cite{Demsar2006} with p < .001 ($\chi^3$(2)=25.09 and $\chi^3$(2)=34.9). A post-hoc Nemenyi test \cite{Demsar2006} indicated a difference between tf-idf and both GloVe and Word2Vec, with the average difference and statistical significance shown in Table \ref{tab:avg_recall_difference}. Based on the average difference, there is a practical significance to the tf-idf vectorization over GloVe and Word2Vec. No statistical or practical difference was present between GloVe and Word2Vec for the RCV1-v2 or Jeb Bush Emails corpora. 

These results indicate generally robust recall performance, particularly with the tf-idf vectorization. Except for the Jeb Bush Emails, and the GloVe vectorization of 20Newsgroups, the median recall was 75\% or better. In the more informal corpora of 20Newsgroups and the Jeb Bush Emails, the GloVe and Word2Vec features did not perform as well. However, this may be due to the corpus specificity of tf-idf compared with the off-the-shelf vocabulary of GloVe and Word2Vec. This suggests that generating corpus-specific GloVe and Word2Vec representations may perform better than the default vocabulary. Future research opportunities exist in optimizing the If-Then rule generation for the tf-idf vectorization and presenting textual and graphical explanations of Word2Vec and GloVe vectorizations. Additionally, exploring corpus-specific versions of Word2Vec and GloVe may bring recall in line with tf-idf, presenting a more efficient yet equally robust option. 

While If-Then rules and graphical representations are acknowledged methods of explainability, there are no agreed-upon quantitative metrics for the explainable artificial intelligence space generally \cite{ArrietaEtAl2020}; in addition, there are also no qualitative or quantitative user studies of the existing prior attempts at explainability in e-discovery TAR \cite{MahoneyEtAl2019, ChhatwalEtAl2018}. Therefore, this represents another likely productive area of future work.

\textbf{Conclusion}: 
This foundational research provides additional support for using the Fuzzy ARTMAP neural network as a classification algorithm in the TAR domain. 
While research opportunities exist to improve recall performance and explanation, the robust recall results from this study and the proof-of-concept demonstration of If-Then rules for tf-idf vectorization strongly substantiate that a Fuzzy ARTMAP-based TAR system is a potentially viable explainable alternative to "black box" TAR systems.


\begin{table}[t]
  \caption{Excerpt of Rule Output for pc.hardware}
  \label{tab:rule_example}
  \centering
  \begin{tabular}{ll}
        \midrule
        \multicolumn{2}{l}{Document is Relevant}\\
        IF& \textit{advance} is \textbf{rarely} prevalent in document\\
        and& \textit{apr} is \textbf{rarely} prevalent in document\\
        and& \textit{bogus} is \textbf{rarely} prevalent in document\\
        and& \textit{browning} is \textbf{highly} prevalent in document\\
        and& \textit{calstate} is \textbf{rarely} prevalent in document\\
        and& \textit{drive} is \textbf{somewhat} prevalent in document\\      
        ...\\
        \midrule
    \end{tabular}
\end{table}

\begin{table}[t]
    \centering
    \caption{Median Metrics by Corpus-Vectorizer}
    \label{tab:median_metrics}
    \begin{tabular}{lllll}
        \toprule
            Corpus & Vectorizer & Recall & Precision & F$_{1}$ \\ 
        \midrule
        20 Newsgroups & GloVe & 0.57 & 0.522 & 0.434 \\ 
        ~ & Word2Vec & 0.772 & 0.41 & 0.523 \\ 
        ~ & tf-idf & 0.94 & 0.367 & 0.53 \\ 
        Jeb Bush Emails & GloVe & 0.622 & 0.07 & 0.125 \\ 
        ~ & Word2Vec & 0.593 & 0.055 & 0.098 \\ 
        RCV1-v2 & GloVe & 0.764 & 0.211 & 0.324 \\ 
        ~ & Word2Vec & 0.752 & 0.187 & 0.292 \\ 
        Reuters-21578 & GloVe & 0.909 & 0.384 & 0.526 \\ 
        ~ & Word2Vec & 0.931 & 0.514 & 0.624 \\ 
        ~ & tf-idf & 0.92 & 0.733 & 0.759 \\ 
        \bottomrule
    \end{tabular}
\end{table}

\begin{table}[t]
    \begin{threeparttable}
        \caption{Average Recall Difference}
        \label{tab:avg_recall_difference}
        \centering
        \begin{tabular}{lll}
            \toprule
                ~ & Reuters-21578 & 20Newsgroups \\
            \midrule
            tf-idf-GloVe & 0.085\tnote{**} & 0.451\tnote{**} \\ 
            tf-idf-Word2Vec & 0.069\tnote{*} & 0.171\tnote{**} \\ 
        \end{tabular}
     \begin{tablenotes}
      \item *p < .05, **p < .01
     \end{tablenotes}
  \end{threeparttable}
\end{table}

\bibliographystyle{ACM-Reference-Format}
\bibliography{legalir-2023-ext-abstract}

\end{document}